\documentclass[fleqn,usenatbib]{mnras}
\usepackage[T1]{fontenc}
\usepackage{ae,aecompl}

\usepackage{graphicx}	
\usepackage{amsmath}	
\usepackage{amssymb}	

\def\la{\raise.5ex\hbox{$<$}\kern-.8em\lower 1mm\hbox{$\sim$}}
\def\ga{\raise.5ex\hbox{$>$}\kern-.8em\lower 1mm\hbox{$\sim$}}
\def\be{\begin{equation}}
\def\ee{\end{equation}}
\def\ba{\begin{eqnarray}}
\def\ea{\end{eqnarray}}

\def\Omegastar{\Omega_\ast}

\def\Omegastardot{\dot{\Omega}_{\ast}}
\def\Mdotin{\dot{M}_{\mathrm{in}}}

\def\Edot{\dot{E}}
\def\Pdot{\dot{P}}
\def\Pddot{\ddot{P}}
\def\Msun{\mathrm{M}_{\sun}}

\def\rin{r_{\mathrm{in}}}
\def\rlc{r_{\mathrm{LC}}}
\def\rout{r_{\mathrm{out}}}
\def\rco{r_{\mathrm{co}}}

\def\Lx{L_{\mathrm{x}}}

\def\Md{M_{\mathrm{d}}}
\def\cs{c_{\mathrm{s}}}

\def\rA{r_{\mathrm{A}}}

\def\Tp{T_{\mathrm{p}}}

\def\dM*{\delta M_*}

\def\Firr{F_{\mathrm{irr}}}

\def\P0min{P_{0,{\mathrm{min}}}}

\def\Alfven{Alfv$\acute{e}$n}
\def\418{SGR 0418+5729}
\def\142{AXP 0142+61}

\def\rbb1{R_{\mathrm{BB1}}}
\def\rbb2{R_{\mathrm{BB2}}}

\title[On the evolution of high-B radio pulsars]{On the evolution of high-B radio pulsars with measured braking indices}

\author[Benli \& Ertan]{
O. Benli\thanks{E-mail:onurbenli@sabanciuniv.edu} and \"{U}. Ertan
\\
Sabanc\i\ University, 34956, Orhanl\i\, Tuzla, \.Istanbul, Turkey
}

\date{Accepted XXX. Received YYY; in original form ZZZ}

\pubyear{----}

\begin{document}
\label{firstpage}
\pagerange{\pageref{firstpage}--\pageref{lastpage}}
\maketitle

\begin{abstract}

We have investigated the long-term evolutions of the high-magnetic field radio pulsars (HBRPs) with measured braking indices in the same model that was applied earlier to individual anomalous X-ray pulsars (AXPs), soft gamma repeaters (SGRs) and dim isolated neutron stars (XDINs). We have shown that the rotational properties (period, period derivative and braking index) and the X-ray luminosity of individual HBRPs can be acquired simultaneously by the neutron stars evolving with fallback discs. The model sources reach the observed properties of HBRPs in the propeller phases, when pulsed radio emission is allowed, at ages consistent with the estimated ages of the supernova remnants of the sources. Our results indicate that the strength of magnetic dipole fields of HBRPs are comparable to and even greater than those of AXP/SGRs and XDINs, but still one or two orders of magnitude smaller than the values inferred from the magnetic dipole torque formula. The possible evolutionary paths of the sources imply that they will lose their seemingly HBRP property after about a few $10^4$ yr, because either their rapidly decreasing period derivatives will lead them into the normal radio pulsar population, or they will evolve into the accretion phase switching off the radio pulses.    

\end{abstract}




\section{INTRODUCTION}

Physical properties of anomalous X-ray pulsars (AXPs), soft gamma repeaters (SGRs), and dim isolated neutron stars (XDINs) indicated by the observations, significantly depend on the actual torque mechanism slowing down these sources. There are two basic torque models that try to explain the evolution of the rotational properties of these young neutron star systems consistently with their X-ray luminosities~: (1) magnetic dipole torque acting on the neutron star rotating in vacuum, (2) the torque of a fallback disc acting on the star through interaction with the magnetic dipole field of the star. The dipole field strength of the star could be overestimated by one or two orders of magnitude with the dipole torque assumption if the star is actually evolving with a fallback disc.

Using the basic principles of the fallback disc model (\citealt{Chatterjee_etal_00, Alpar_01}), we have developed a long-term evolution model considering the inactivation of the disc below a critical temperature, the contribution of the cooling luminosity of the star to X-ray irradiation, and the effect of the X-ray irradiation on the evolution of the disc. We applied the same evolution model earlier to individual AXP/SGRs including the so-called low-B magnetars (\citealt{Alpar_etal_11, Benli_etal_13, Benli_Ertan_16}), to the six XDINs with confirmed period and period derivatives \citep{Ertan_etal_14}, and to a high--B radio pulsar (HBRP) with an anomalous braking index ($n \simeq 1$), namely PSR J1734--3333 \citep{Caliskan_etal_13}. These model applications are self-consistent in that the results are obtained with similar basic disc parameters that are expected to be similar for the fallback discs of different systems.

In this work, we use the same model to investigate the properties of the 3 HBRPs with measured braking indices. Since the model is described in the earlier works (see e.g. \citealt{Caliskan_etal_13, Ertan_etal_14, Benli_Ertan_16}), we do not discuss the model details here. In Section 2, we briefly explain the model parameters and the evolution of a neutron star with a fallback disc. In Section 3, we give the observational properties of the 3 HBRPs. Our results are given in Section 4. We summarize our conclusion in Section 5.

\section{EVOLUTION OF A NEUTRON STAR WITH A FALLBACK DISC}

Evolutionary phases of a neutron star evolving with a fallback disc depend on the initial conditions: the magnetic dipole field strength on the pole of the star, $B_0$, the initial period, $P_0$, and the disc mass, $\Md$. We assume that the disc is initially extended to a radius $\rout$ at which the effective temperature decreases to a critical value, $\Tp$, the minimum temperature that can keep the disc matter in a viscously active state. During the long-term evolution, $\rout$ gradually propagates inwards with decreasing X-ray irradiation flux that can be written as $\Firr \simeq 1.2~C~\Lx/\pi r^2$ where $C$ is the irradiation efficiency parameter that depends on the disc geometry and albedo of the disc surface and $\Lx$ is the X-ray luminosity of the neutron star \citep{Fukue_92}. The values of $C$ in the $(1-7) \times 10^{-4}$ range can produce the optical and infrared emission spectra of AXP/SGRs \citep{Ertan_Caliskan_06, Ertan_etal_07b}. Keeping $C$ in this range, the general X-ray luminosity and the rotational properties of AXP/SGRs can be obtained with $\Tp \sim 100$ K \citep{Ertan_etal_09}. We solve the disc diffusion equation using the kinematic viscousity $\nu = \alpha \cs h$, where $\cs$ is the sound speed, $h$ is the pressure scale-height of the disc. We use the same $\alpha$ parameter ($\alpha = 0.045$) that was used earlier for AXP/SGR and XDINs. 

In the accretion phase, to find the total torque acting on the star we integrate the magnetic torque between the conventional \Alfven~radius, $\rA \cong (G M)^{-1/7}~\mu^{4/7} \Mdotin^{-2/7}$ \citep{Davidson_Ostriker_73, Lamb_etal_73}, and the co-rotation radius, $\rco = (G M / \Omegastar^2)^{1/3}$ where $M$, $\Omegastar$ and $\mu$ are the mass, the angular frequency and the magnetic dipole moment of the neutron star,  $G$ is the gravitational constant and $\Mdotin$ is the rate of mass-flow to the inner disc. The integrated spin-down torque can be written as, $N = I ~\Omegastardot \simeq \frac{1}{2} \Mdotin ~(G M \rin)^{1/2} ~(1 - (\rin/\rco)^3)$ (\citealt{Ertan_Erkut_08}). The radius $\rin$ can be considered as the inner radius of the thin disk, while the boundary region of interaction between the field lines and the inner disk extends from $\rin$ to $\rco$. For the accretion phase, we set $\rin = \rA$ in the torque expression. Since the critical condition for the transition between the accretion and the propeller phase is not well known, we use the simplified condition $\rA = \rlc$  for the accretion-propeller transition, where $\rlc = c/\Omegastar$ is the light cylinder radius and $c$ is the speed of light. That is, when the calculated $\rA$ is found to be greater than $\rlc$, the system is in the propeller phase. In this phase, we calculate the total torque substituting $ \rin = \rlc$ in the torque equation. 
In the propeller phase, we assume that all the mass flowing to the inner disc is expelled from the system allowing pulsed radio emission. (see \citealt{Ertan_Erkut_08} for the details of the torque model). Previously, \cite{Chen_Li_16} also studied the long-term evolution of PSR J1734--3333 using a different torque model and could produce the anomalous braking index of the source. Nevertheless, in their model, the rotational properties of the source are  
produced with disk luminosities that are well above the observed luminosities. This discrepancy between our results and those found by \cite{Chen_Li_16} is mainly due to the differences in the torque calculations. In this work, we use the same relatively efficient disk torque model that can also reproduce the X-ray luminosities and the rotational properties of different sources from different populations in a self-consistent way.  

In line with the long-term evolution models of AXP/SGRs and XDINs with fallback discs, in this work we also take $\alpha =0.045$ and $\Tp = 100$~K and $C$ in the $(1-7) \times 10^{-4}$ range for all the sources. To test the model, we repeat the simulations tracing the initial conditions, $P_0$, $B_0$ and $\Md$. We count a model as a reasonable alternative representation of the evolution of a given HBRP, if it reproduces the X-ray luminosity, $\Lx$, the period, $P$, the period derivative, $\Pdot$, and the second period derivative, $\Pddot$, (or the breaking index, $n = 2 - P \Pddot / \Pdot^2$) simultaneously (For the details of the model see e.g. \citealt{Benli_Ertan_16}).

\section{SOURCE PROPERTIES}

{\bf PSR J1119--6127} was discovered in the Parkes Multibeam Pulsar Survey with a rotational period $P = 0.408$~s and a period derivative $\Pdot = 4.02 \times 10^{-12}$~s~s$^{-1}$. The braking index $n = 2.91 \pm 0.05$ was measured using a 1.2 year data \citep{Camilo_etal_00}. Using more than 12 year timing data, \cite{Weltevrede_etal_11} obtained $n = 2.684 \pm 0.002$. A more recent analysis covering 16 year data excluding the imprint of glitch activities yielded $n \simeq 2.7$ \citep{Antonopoulou_etal_15}. The source has a rotational power $\Edot = 2.3 \times 10^{36}$ erg s$^{-1}$. \cite{Caswell_etal_04} estimated a distance of 8.4~kpc using neutral hydrogen absorption measurement to the supernova remnant (SNR) G292.2--0.5 and estimated an upper limit to the SNR age around 10 kyr. Through spectral fit to the X-ray spectra, \cite{Ng_etal_12} estimated the bolometric X-ray luminosity as $\Lx = (1.1-3.8) \times 10^{33}$~erg~s$^{-1}$ in the 0.5--7~keV band, assuming $d = 8.4$~kpc. 

{\bf PSR J1734--3333}~ has period $P = 1.17$ s, period derivative $\Pdot = 2.28 \times 10^{-12}$ s s$^{-1}$ and period second derivative $\Pddot = 5.3 \times 10^{-24}$ s s$^{-2}$ ($n = 0.9 \pm 0.2$, \citealt{Espinoza_etal_11}). This braking index is the lowest among the young radio pulsars. The range of bolometric X-ray luminosity of the source corresponding 25\% uncertainty in the distance estimate is $7.3 \times 10^{31} - 6.6 \times 10^{32}$ erg s$^{-1}$ \citep{Olausen_etal_13}. The age of the SNR associated with the source is estimated to be greater than 1300 yr  \citep{Ho_Anderson_12}. 

{\bf PSR B1509--58} was discovered in the soft X-ray band with \textit{Einstein Observatory} \citep{Seward_Harnden_82} and subsequently observed also in the radio band \citep{Manchester_etal_82}. It has a period $P \simeq 0.15$ s and a period derivative $\Pdot \simeq 1.5 \times 10^{-12}$~s s$^{-1}$, which give $\Edot \simeq 1.7 \times 10^{37}$~erg~s$^{-1}$ and a characteristic age of 1570 yr. The age of SNR G320.4--01.2 associated with the source is estimated to be less than about $ 1700$~yr \citep{Gaensler_etal_99}. Recent analysis of the off-pulsed X-ray spectrum gives the $0.5-7$~keV X-ray luminosity between $10^{33}$~erg~s$^{-1}$ and $2 \times 10^{34}$~erg~s$^{-1}$ for the distance $ d = 5.2$~kpc \citep{Hu_etal_17}. The persistent braking index of the source is measured to be $n \simeq 2.84$ \citep{Kaspi_etal_94}. 

\section{RESULTS AND DISCUSSION}

Tracing the initial conditions $B_0$, $P_0$ and $\Md$, we have tested whether our long-term evolution model can produce the observed rotational properties and X-ray luminosities of the three HBRPs. Taking the viscousity parameter $\alpha = 0.045$, the critical temperature $\Tp = 100$~K and keeping the irradiation parameter $C$ in the $(1-7) \times 10^{-4}$ range, like in the models applied to AXP/SGRs and XDINs, we have found that the model can reproduce the properties of the 3 HBRPs for certain ranges of the initial conditions. The evolution of PSR J1734--3333 was studied earlier by \cite{Caliskan_etal_13} with $\alpha = 0.03$. For a complete analysis of HBRPs with measured braking indices, we have re-analysed the source with $\alpha = 0.045$, and found similar evolutionary curves to those found by \cite{Caliskan_etal_13}.

It is seen in Figs (1--3) that the model can reproduce the individual source properties ($\Lx$, $P$, $\Pdot$ and the braking index $n$) simultaneously for PSR J1119--6127, PSR J1734--3333 and PSR B1509--58. The model parameters are given in the figures. The three HBRPs are associated with SNRs. The ages of the sources indicated by the evolutionary curves are consistent with the estimated ages of their SNRs (see Section 3). The estimated ages and the observed properties of the sources are listed in Table \ref{tab:properties}.

\begin{figure}
\centering
\includegraphics[width=\columnwidth,angle=0]{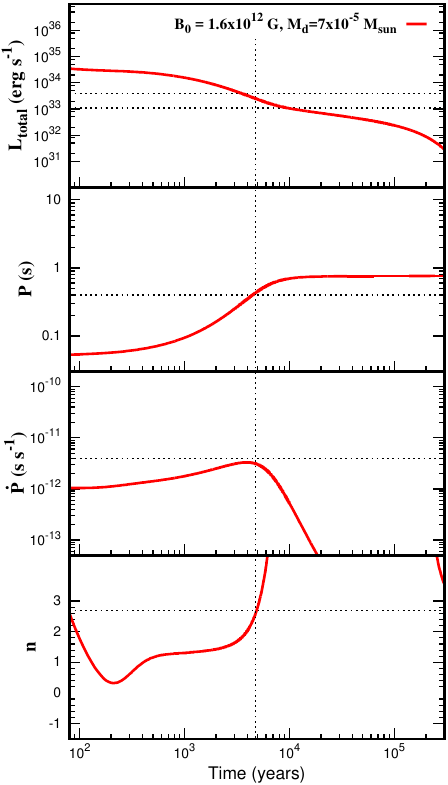}
\caption{Illustrative model curve for PSR J1119--6127. The properties of the source are produced at age $\sim 5 \times 10^{3}$ yr (the vertical dashed line). From the top to the bottom panel, the model curves represent $\Lx$, $P$, $\Pdot$ and $n$ evolutions. For this model, $P_0 = 50$ ms, $C = 1 \times 10^{-4}$ and $\Tp = 100$ K. The $\Md$ and $B_0$ values are given in the top panel. The horizontal dashed lines show the observed properties. The double dashed lines in the top panel show the uncertainty range of the X-ray luminosity. }
\label{fig:1119}
\end{figure}
\begin{figure}
\centering
\includegraphics[width=\columnwidth,angle=0]{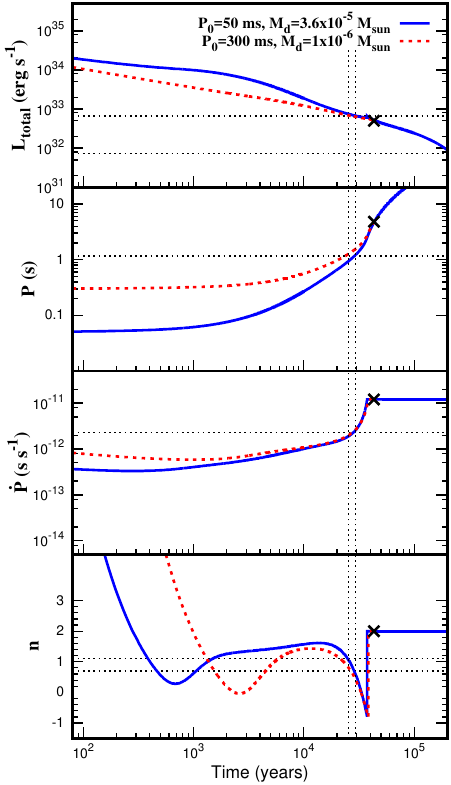}
\caption{Illustrative model curves for PSR J1734--3333. For these models, $B_0 = 2 \times 10^{12}$~G, $\Tp = 100$ K and $C = 7 \times 10^{-4}$.  In these two models, we use the $P_0$ and $\Md$ values given in the top panel. The cross signs indicate the times at which the sources intersect the pulsar death-line. 
}
\label{fig:1734}
\end{figure}
\begin{figure}
\centering
\includegraphics[width=\columnwidth,angle=0]{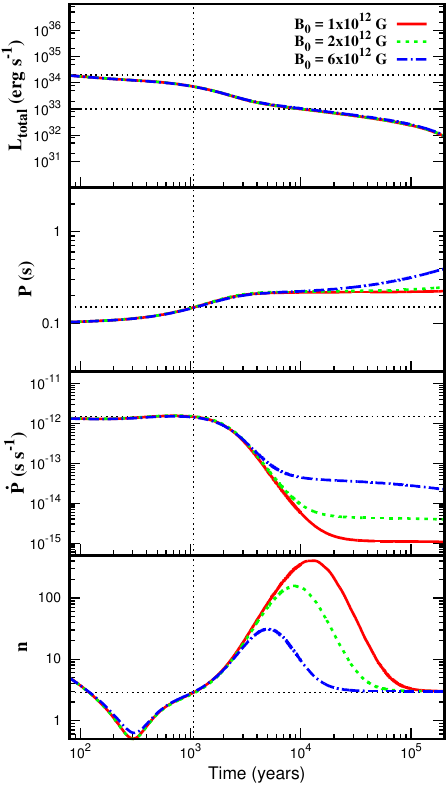}
\caption{Model curves for PSR B1509--58. For these three illustrative models, we take $\Md \sim 2 \times 10^{-5} \Msun$, $P_0 = 100$ ms, $C = 1 \times 10^{-4}$ and $\Tp = 100$ K.  The dotted, dashed and solid model curves are obtained with the $B_0$ values given in the top panel.
}
\label{fig:B1509}
\end{figure}

Unlike for AXP/SGRs and XDINs, our model does not significantly constrain the dipole field strength, $B_0$, on the pole of the star. Similar evolutionary curves can be obtained with $B_0$ values from $\sim 10^{12}$~G to $\sim 10^{13}$~G. The results of earlier work indicate that XDINs have relatively weak fields with  $10^{11}$~G $\lesssim B_0 \lesssim 10^{12}$~G, while for AXP/SGRs $B_0 \gtrsim 10^{12}$~G \citep{Ertan_etal_14, Benli_Ertan_16}. This imply that HBRPs could have field strengths similar to or possibly greater than those of  AXP/SGRs in the fallback disc model. For these three sources, the radio pulsar death line does not impose a lower limit on $B_0$ either, due to their short periods ($\lesssim 1$ s).

For PSR J1119--6127 and PSR B1509--58, we obtain the best results with the $P_0$ and $\Md$ values given in Figs. 1 and 3. Nevertheless, reasonable model curves could be obtained with large range of these initial parameters. For PSR J1734--3333, the evolution is not very sensitive to the initial period either. The properties of this source could be produced with $P_0$ and  $\Md $ values in the ranges  $ \sim 50-300$ ms and $\sim 10^{-6}-10^{-5} ~\Msun$  respectively (see Fig. \ref{fig:1734}).  The disc masses of the three HBRPs, like their field strengths, seem to be similar to or greater than those of AXP/SGRs. Our results imply that  these three sources are not likely to be identified as HBRPs after ages of  a few $10^4$ yr (see Figs. 1--3) depending on their actual initial disc masses. One of the sources (PSR J1734--3333) is evolving into the AXP phase, while the others are likely to continue their evolution as radio pulsars, but with rapidly decreasing $\Pdot$ to the normal radio pulsar properties. A more detailed analysis is required to determine the entire domain of the initial conditions leading a neutron star to HBRP properties and the duration of these phases with relatively high $\Pdot$ values. 

In the accretion phase, the sources cannot emit radio pulses. For the self-consistency of the model, the observed properties of HBRPs should be acquired in a phase that allows radio pulses, that is, in the propeller phase. Our results for the three sources satisfy this requirement as well. A radio pulsar with a fallback disc can evolve into accretion phase depending on the initial conditions (see e.g. Fig. \ref{fig:1734}). Once a source enter this phase, it is not likely to evolve into the radio phase. Because, in most cases, the accretion phase terminates with a rotational rate that is not sufficient to power pulsed radio emission. For instance, in this model, XDINs slowed down in the accretion phase in the past, and are currently in the propeller phase. During the accretion-propeller transition, all these sources have periods that place them below the pulsar death-line in the $B_0$-$P$ plane (see Fig. 4. in \citealt{Ertan_etal_14}).    

The differences of our model from the earlier fallback disc models \citep{Chatterjee_etal_00, Alpar_01, Menou_etal_01} was discussed in \cite{Caliskan_etal_13}. The basic differences are: (1) the inactivation temperature of the disc in our model ($\sim 100$~K) which is much lower than in the other models, (2) the torque calculation, and more importantly (3) the condition for the transition between the propeller and the accretion phases. In our model, the sources are accreting matter from the disc over a large range of accretion rates in the spin-down phase. Observations of the transitional millisecond pulsars that show transitions between the X-ray pulsar and the radio pulsar phases at very low X-ray luminosities \citep{Archibald_etal_15b, Papitto_etal_15} seem to be consistent with our simplified condition for the onset of the propeller phase. Recently, \cite{Ertan_17} estimated the critical accretion rate for this transition depending on the period and the dipole field strength of the star. These critical accretion rates estimated by simple analytical calculations are in agreement with the rates obtained by our simplified condition ($\rA = \rlc$) for the accretion-propeller transition. 

\begin{table*}
\centering
\caption{The observed properties of high-magnetic field radio pulsars and their ages found from the model}
\label{tab:properties}
	\begin{tabular}{lccccr} 
	\hline
		Name			 			& $P$ (s) & $\Pdot$ (s s$^{-1}$)        & n         			     & $\Lx$ (erg s$^{-1}$)   		 & Age (yr) \\ 
		\hline				 
		PSR J1734-3333     & 1.17	  	& $2.28 \times 10^{-12}$    & $0.9 \pm 0.2$	 & $7.3-66 \times 10^{31}$ 	 &  $2.5-3 \times 10^4$  \\
		PSR J1119--6127		& 0.41   	& $4.02 \times 10^{-12}$    &    2.7		   		 & $1.1-3.8 \times 10^{33}$ &  $5 \times 10^{3}$   \\ 
		PSR B1509--58		& 0.15   	& $1.53 \times 10^{-12}$    &    2.84	   			 & $1-20 \times 10^{33}$		 & $ 1 \times 10^{3}$	\\
	\hline
	\end{tabular}
\end{table*}

\section{CONCLUSION}

We have investigated the long-term evolution of the three high--B radio pulsars (HBRPs) with measured braking indices, namely PSR J1734--3333, PSR J1119--6127 and PSR B1509--58, using the same model that was applied earlier to AXP/SGRs and XDINs. We have shown that the neutron stars starting their evolution from a certain domain of initial conditions ($B_0$, $P_0$, $\Md$) can reach the observed X-ray luminosity, period, period derivative and braking index of each of these HBRPs simultaneously through evolutions with fallback discs and with conventional magnetic dipole fields. For all these sources, the model reproduces the observed properties at ages that are in good agreement with the estimated ages of the supernova remnants associated with these systems. In the model, the three HBRPs are currently in the propeller phase when the accretion on to the star is not allowed, which is consistent with the radio pulsar property of these sources.

We obtain the properties of the 3 HBRPs with relatively high disc masses in comparison with those of AXP/SGRs and XDINs. In the fallback disc model, our results indicate that the dipole fields of HBRPs could be in the $\sim 10^{12}$~G to $\sim 10^{13}$~G range that is similar to AXP/SGR field range. This means that a fraction of the HBRPs could have evolutionary connections with AXP/SGRs (see e.g. the model curves for PSR J1734--3333 with $B_0 \simeq 2 \times 10^{12}$~G), while the remaining fraction could evolve as radio pulsars over their observable life-times (see the illustrative model curves in Figs \ref{fig:1119} and \ref{fig:B1509}). 

Our results, comparing with the results of earlier work, also show that the individual source properties of AXPs, SGRs, XDINs and HBRPs can be re-produced in the same model as a natural outcome of evolutions of neutron stars with fallback discs. It is the differences in the three initial conditions ($B_0$, $P_0$, $\Md$) that lead to emergence of observed diversity of these neutron star populations. A detailed analysis of the evolutionary connections between different isolated neutron star populations in the fallback disc model will be studied in an independent paper. In particular for the 3 sources studied in this work, our results imply that they will lose their HBRP property, inferred from the observed $P$ and $\Pdot$ with purely magnetic dipole torque assumption, before the ages of a few $10^4$ yr. In the subsequent evolutionary phases, they could be identified either as normal radio pulsars because of their rapidly decreasing $\Pdot$ (see Figs. \ref{fig:1119} and \ref{fig:B1509}) or they could enter the accretion phase switching off their radio pulses (see Fig. \ref{fig:1734}) and possibly switching on the AXP/SGR properties. 

\section*{Acknowledgements}

We acknowledge research support from Sabanc{\i} University, and from T\"{U}B\.{I}TAK (The Scientific and Technological Research Council of Turkey) through grant 116F336.



\bibliographystyle{mn2e}
\bibliography{benli} 





\bsp	
\label{lastpage}
\end{document}